\documentstyle[preprint,aps]{revtex}
\begin{document}
\draft
\preprint{$
\begin{array}{l}
\mbox{LBL-37189}\\ [-3mm]
\mbox{UCB-PTH-95/12}
\end{array}$}

\title{Isospin Coherence and
Final-State Scattering of \\Disoriented Chiral
Condensate
\thanks{
\baselineskip=12pt
This work was supported by the Director, Office of Energy
Research, Office of High Energy and Nuclear Physics, Division of High
Energy Physics of the U.S. Department of Energy under Contract
DE-AC03-76SF00098.}
}
\author{Zheng Huang
and Mahiko Suzuki}

\address{{
Theoretical Physics Group,
Lawrence Berkeley Laboratory\\
\baselineskip=12pt       University of California,
    Berkeley, California 94720, USA}}

\date{April 5, 1995}
\maketitle
\begin{abstract}
We examine the validity of the notion of the coherent
state for pions and
the quantum scattering effect in the final state of pion emission.
When the number of particles is large,
the effect caused by the small but finite mass difference
between the neutral
and charged pions can add up substantially in the quantum evolution of
an initially coherent state.
As a result, the states with
quite different numbers of neutral or charged pions are essentially
{\it incoherent}.
The importance of the quantum scattering in the final-state
isospin charge distribution of a disoriented chiral condensate (DCC)
is investigated.
We find that the scattering effect significantly reduces the spectacular
Centauro and anti-Centauro events. The deformation of a charge
distribution $dP/df$ predicted by the classical field theory
is significant only for a  DCC with the size of 10 fm or more.
\end{abstract}
\newpage
\narrowtext
\section{Introduction}
Recently there has been a growth of interest in the subject
of coherent pion radiation, in conjunction with the conjectured
 disoriented
chiral condensate (DCC) \cite{dcc,bj,num,hsw}.
It is suggested that the very high energy
hadron-hadron or nucleus-nucleus collisions may create an extended region
of space-time containing strong-interaction vacuum with a non-standard chiral
orientation. In the language of the linear $\sigma$-model the order parameter
which usually points in the $\sigma$ direction is presumed to be in some
other direction. In the ``Baked Alaska'' scenario \cite{bj},  the disoriented
vacuum is found in the expanding shell of hot collision debris.
Indeed, the numerical simulations of the classical equation of motion
show some evidence of
growth of  long wavelength modes when the expanding system is out of
equilibrium, leading to the formation of large isospin correlated domains
\cite{num}.
When the hot shell hadronizes and breaks up, the
disorientation is radiated away in Goldstone modes (pions) of a common
isospin. The most natural way to connect the classical wave
with a quantum state is the coherent state formalism in the limit of
a large number
of quanta.
Such coherent pulses of semiclassical pion fields would
lead to anomalously large fluctuations event-to-event in the ratio of neutral
to charged pions produced. Assuming the all cartesian isospin directions are
equally probable, one derives the probability ($P$) distribution \cite{dcc,bj}
\begin{equation}
\frac{dP}{df}=\frac{1}{2\sqrt{f}} \label{frac}
\end{equation}
with the definition of neutral fraction $f=N_{\pi^0}/N$.

In this paper, we examine the validity of the notion of the coherent
state for pions, in particular,
the quantum disturbance to coherence in the final state of pion emission.
We find that the effect caused by the small but finite mass difference
between the neutral
and charged pions adds up substantially in the quantum evolution of
an initially coherent state of pions. As a result, the states with
quite different numbers of neutral or charged pions are essentially
{\it incoherent} due to the different  phase shifts!
This  phenomenon occurs only in the system where the number of isospin
carrying particles (e.g., the pions) are large and the evolution duration
of the system is long compared to O$(1/m_\pi )$. Although the notion of the
isospin coherence for a many-pion system can be entirely inadequate, the
prediction of the neutral pion distribution $P(f)$ of DCC in (\ref{frac}) is
not altered by the phase shift caused only by the mass terms.
We shall compute the modification of (\ref{frac}) due to
the quantum rescattering effects when the interactions are included.

\section{Isospin Coherence of DCC}
The simplest
quantum description
of a DCC state is given in terms of the coherent state of the classical
pion and $\sigma$ fields in the linear $\sigma$ model.
When the DCC of a large isospin correlated domain is created, one considers
an ideal limit where there is no spatial dependence of the DCC
chiral orientation. Since we expect that only small isospin
states are created, we consider for simplicity the $I=0$ state which is
constructed by the superposition of the oriented DCC states denoted
by $|\theta ,\varphi\rangle$ over the isospin direction $(\theta ,
\varphi )$ \cite{kt}:
\begin{equation}
|I=0\rangle_{\rm DCC}=\int d\Omega
(\theta ,\varphi )|\theta ,\varphi \rangle . \label{1}
\end{equation}
The DCC state relaxes to the true vacuum state by radiating the Goldstone
modes. Note that the light particles emitted  during this period are
not the Goldstone modes associated with the true
vacuum, but the disoriented vacua,  which are in general linear combinations of
$\pi$ and $\sigma$ (they are ``disoriented'' Glodstone bosons).
The classical $\sigma$ field evolves as well
as the pion field, eventually the $|I=0\rangle_{\rm DCC}$ state
turns into the true vacuum (a constant $\sigma$ field) plus many outgoing
low-energy pions.

If one assumes that all pions radiated from a DCC during the relaxation
end up in the same orbital state in the true vacuum,
 the state to which
$|I=0\rangle_{\rm DCC}$ relaxes can be written in a unique, simple form
\cite{hs,kt}. The projection of this state
onto an eigenstate of the pion number
operator $\hat{N}=\hat{N}_++\hat{N}_-+\hat{N}_0$ is
\begin{equation}
|I=0;N\rangle=\frac{1}{\sqrt{(N+1)!}}
(2a^\dagger_+a^\dagger_--a^\dagger_0
a^\dagger_0)^{N/2}|0\rangle , \label{2}
\end{equation}
where $|0\rangle$ is the true vacuum in which $\langle \sigma\rangle =f_\pi$,
and
${ a}_\pm^\dagger$ and $a_0^\dagger$
 are the pion creation
operators for $\pi^\pm$ and $\pi^0$.
The same neutral pion distribution $P(f)$ as in (\ref{frac})
is obtained by expanding (\ref{2}) in
the basis of definite charge states $|N_0\rangle$'s
for $f=N_0/N$ in the large $N$, $N_0$ limit.
 Note that
$N_+=N_-=(N-N_0)/2$ for  $|N_0\rangle$. As long as one accepts the assumption
of  {\em a single orbital wavefunction} for pions, the state
$|I=0;N\rangle$ is not subject to any final-state interaction corrections (up
to an overall phase) since it is the only $I=0$ state made of $N$ pions.

However, such a picture is oversimplified. It is not clear how all final
pions in the true vacuum can retain the shape of the original wavefunction
after a continuous emission of the time-varying
``disoriented'' Goldstone modes. These Goldstone modes emitted in
a disoriented vacuum background field are different from those of the true
vacuum in that they contain the $\sigma$'s as well, which must turn into
pions  when the disoriented background changes into the true
vacuum surrounding the DCC domain. As a result,
 more than one orbital wavefunction must be
allowed for pions and many other isosinglet $N$ pion states can be
constructed. Furthermore, the quantum scattering causes the transitions
among those $I=0$ states, and the coherence among different $|N_0\rangle$
states may be lost.

Even if all final pions end up in the same orbital configuration, the isospin
breaking effects will break up the coherence among $|N_0\rangle$'s.
The isospin breaking caused by the electromagnetic interactions and the
$u$- and $d$- quark mass difference is usually ignored in the DCC state and in
its time development, which is a good approximation in most cases. However, it
can be a problem when one is concerned with the coherence among states
$|N_0\rangle$ with
different $N_0$'s in $|I=0;N\rangle$. The $\pi^\pm$ and
$\pi^0$ actually oscillate with slightly different frequencies because of their
electromagnetic mass difference $\Delta m_\pi=m(\pi^\pm )-m(\pi^0)=4.6$ MeV.
Consider the single momentum mode ${\bf k}=0$.
$a^\dagger_+a^\dagger_-|0\rangle$
and the terms $a^\dagger_0a^\dagger_0|0\rangle$
acquire a relative phase $\delta (t)=2\Delta m_\pi t$ due to
$\Delta m_\pi$, so that (up to an overall phase)
\begin{equation}
|t;N\rangle= \frac{1}{\sqrt{(N+1)!}}
(2a^\dagger_+a^\dagger_--e^{i\delta (t)}a^\dagger_0
a^\dagger_0)^{N/2}|0\rangle . \label{4}
\end{equation}
At $t=4$ fm/c,
$\delta (t)\sim 0.2$. Although the phase for each $\pi^0$ pair is small,
it accumulates with $N_0$.
The relative phase between the two charge states
 $|N_0\rangle$ and
$|N_0+\Delta N_0\rangle$, differing in the number of neutral
pions by $\Delta N_0$, gets amplified by $\Delta N_0$ times
and becomes  $\delta (t)\Delta N_0$.
When $N$ is let to infinity with a fixed
$\Delta f=\Delta N_0/N$, this phase goes to infinity
and wipes out any coherence between the two states. For finite
$N$ and $\Delta N_0$, the initial coherence between
$|N_0\rangle$ and
$|N_0+\Delta N_0\rangle$ is lost in a characteristic time scale
$\tau \sim 1/(\Delta N_0\Delta m_\pi)$.
For $\Delta f=0.2$  and $N=50$, for instance, the isospin breaking
effect rapidly accumulates in $\tau \simeq 4$ fm/c, leaving only
the $I_3$ as a good quantum number. When the mutual coherence among
different $|N_0\rangle$ states in (\ref{4}) is lost,
$\pi^\pm$ and $\pi^0$ thereafter behave without the overall isosinglet
constraint. However, it is interesting to note that
the phase $\delta (t)$
does not affect
the classical prediction
(\ref{frac}) on the distribution $P(f)$ \cite{ak}.
Our picture  is very close to a purely classical description where
the states with definite (classical) isospin orientation are produced
incoherently, and the charge ratio
$N_+/N$, $N_-/N$ and $N_0/N$ are identified with the isospin direction of
the classical pion field.

\section{Quantum Rescattering Effects}
We have argued that after the relaxation of the DCC into
the true vacuum, the different charge states $|N_0\rangle$'s should
be treated as mutually incoherent.
As we shall study
how the final state quantum scattering changes the prediction in (\ref{frac}),
we must start with a pion-number eigenstate $|N_0\rangle$ instead of
the isosinglet state in (\ref{2}).
The quantum scattering processes $\pi^0\pi^0\leftrightarrow
\pi^+\pi^-$ cause the transition between $|N_0\rangle$
and $|N_0\pm 2\rangle$, which can distort the distribution $P(f)$.
One needs to worry
if there will be enough quantum scattering of pions to dilute the effect
of interest in (\ref{frac}). Put  more explicitly, suppose a DCC
charge state initially consists
of only $N$ $\pi^0$'s, we would like to calculate
how many $\pi^+$'s or $\pi^-$'s end up in the detector
after the quantum scatterings are included.

The relevant quantity for determining the importance of the quantum scattering
is the mean free path ($l_0$) of a
physical pion in a cloud of emitted pions at the
time (say, $t=0$) when the DCC decays. If the size of the system ($R_D$) at the
time is much smaller than $l_0$, the pions essentially decouple from the system
and propagate freely with little interactions. The opposite limit is
$R_D\gg l_0$, in which case
a pion will encounter many rescatterings before leaving
the system and  the
measured value $f$ ($\simeq 1/3$) has little trace of the
original charge fraction $f_0$. In the general case, we have to conduct a
quantitative computation.
Since the pions from a DCC are soft in general, the mean free
path of collision is much longer than a typical size of DCC. We shall
make the approximation that the emitted pions
stream out almost freely, i.e.\ that
the collision does not slow down
the diffusion of the pion cloud. In this approximation,
 there is a simple relation between
the speed of diffusion ($v_{\rm d}$) and the
average relative velocity ($v_{\rm rel}$) for pion collisions.
With this simplification (an underestimate), we shall show that the quantum
corrections are characterized by a rather simple and intuitive
factor $\exp [-4R_D/(3l_0)]$. If the size of the diffusing cloud is a
significant fraction of the mean free path, the quantum rescattering is
non-negligible and will dilute the Centauro and anti-Centauro
events \cite{cen}.

We are concerned with the evolution of the charge composition of the
pion cloud with three charge states denoted by $0, +, -$.
For simplicity we assume
that the pion density is spatially uniform inside the DCC and the typical
momentum is determined by the uncertainty principle
$\langle   p\rangle \approx 1/R_D$.  For the DCC of an uniformly aligned
orientation, this is a good estimate, but it is an underestimate for the
DCC having domain structure.
When the typical energy of pions is low, the
relevant reaction is the two-body
$\pi\pi$ scattering. The neutral pion state, for
instance, is depleted by the annihilation process
$\pi^0\pi^0\rightarrow\pi^+\pi^-$ and replenished by the reverse process
$\pi^+\pi^-\rightarrow\pi^0\pi^0$.
Under these assumptions, the time evolution of the numbers of three charge
particles can be readily derived from the Boltzmann transport equation:
\begin{eqnarray}
\frac{dN_0}{dt} & = &
-\frac{2\langle   v_{\rm rel}\sigma_{\pi\pi}\rangle}{V(t)}
 N_0^2+\frac{2\langle   v_{\rm rel}\sigma_{\pi\pi}\rangle}{V(t)}N_+N_-,\\
\frac{dN_+}{dt}& = & \frac{\langle   v_{\rm rel}\sigma_{\pi\pi}\rangle}{V(t)}
 N_0^2-\frac{\langle   v_{\rm rel}\sigma_{\pi\pi}\rangle}{V(t)}N_+N_-,\\
\frac{dN_-}{dt}& =& \frac{\langle   v_{\rm rel}\sigma_{\pi\pi}\rangle}{V(t)}
 N_0^2-\frac{\langle   v_{\rm rel}\sigma_{\pi\pi}\rangle}{V(t)}N_+N_-,
\end{eqnarray}
where the cross section $\sigma_{\pi\pi}$ is defined to be that of $\pi^+\pi^-
\rightarrow \pi^0\pi^0$ (note that $\sigma (\pi^+\pi^-
\rightarrow \pi^0\pi^0)=
\frac{1}{2}\sigma(\pi^0\pi^0\rightarrow\pi^+\pi^-)$). $V(t)$ is the
volume of pion cloud enclosed by the outgoing pion front.
$\langle   v_{\rm rel}\sigma_{\pi\pi}\rangle$
is the average value of the relative velocity
times the cross section.
Adding the above
three equations  gives the total number conservation $dN/dt=0$ where
$N = N_0(t)+N_+(t)+N_-(t)$, which is the consequence of the
fact that only two-body
reactions are considered in collision terms in the Boltzmann equation.
Since initially $N_+(0)=N_-(0)$, the evolution
equations guarantee that this remains true at later times.
In this case, we obtain a simple analytic solution. Define
$f_0\equiv {N_0(0)}/{N}$ and $f(t)\equiv {N_0(t)}/{N}$.
The solution is
\begin{equation}
f(t)=\frac{(f_0+1)+(3f_0-1)e^{-{\cal F}(t)}}{3(f_0+1)-(3f_0-1)e^{-{\cal
F}(t)}},
\label{f}
\end{equation}
where ${\cal F}(t)$ determines
the time dependence of charge composition due to the
final-state scatterings,
\begin{equation}
{\cal F}(t) =
2N\int_{0}^{t}\frac{\langle   v_{\rm rel}\sigma_{\pi\pi}\rangle}{V(t)}dt.
\label{in}
\end{equation}

Before we estimate the ${\cal F}(t)$,
some remarks on the general features of (\ref{f}) are in order.
${\cal F}(t)$ is a positive number and $e^{-{\cal F}(t)}<1$.
Generally $f$ depends
on ${\cal F}(t)$ {\it and} the initial value $f_0$. It is easy to verify that
when $f_0<1/3$, $f>f_0$ and when $f_0>1/3$, $f<f_0$ at later times ($t>0$).
In other words, if the system initially has fewer $\pi^0$'s compared to
$\pi^+$'s or $\pi^-$'s, the quantum scattering tends to increase the number of
$\pi^0$'s; if more $\pi^0$'s, the quantum scattering tends to decrease it.
$f_0=1/3$  is a special value, in which case the time
dependence of $f$ completely disappears and $f$ keeps a constant $f=f_0=1/3$.
In the classical theory, $f_0$ ranges from $0$ to $1$ and the end points
correspond to the Centauro and anti-Centauro events.  These
extreme situations do not occur in the quantum theory. In fact, $f$ can never
be equal to $0$ or $1$ as long as ${\cal F}(t)\neq 0$.
Since the state of all
$\pi^0$ or all $\pi^{\pm}$ is  degraded by the charge conversion processes,
the quantum rescattering
squeezes the allowed range for $f$ from the both ends.
The upper and lower limits on $f$ are:
\begin{equation}
f_{\rm max} =\frac{1+e^{-{\cal F}}}{3-e^{-{\cal F}}}\quad ,
\quad f_{\rm min} =\frac{1-e^{-{\cal F}}}{3+e^{-{\cal F}}}.
\end{equation}
The ranges for $f<f_{\rm min}$ and $f>f_{\rm max}$ are
excluded by the quantum
rescattering.
If the rescattering is  very strong,
the factor $e^{-{\cal F}}$ would vanish, and
$f_{\rm max}$ and $f_{\rm min}$ converge to a single point $f=1/3$.
Even for relatively weak
scatterings of interest, the most spectacular Centauro and anti-Centauro
events are diluted to some extent.

We would like to estimate the integral in (\ref{in}) for a realistic
DCC.
We assume that the volume $V(t)$ of the pion cloud is spherical and diffuses
with a constant velocity $v_{\rm d}$ as
\begin{equation}
V(t) = \frac{4\pi}{3}(R_D+v_{\rm d}t)^3. \quad\quad\quad (0<t<\infty)
\end{equation}
The time dependence of the $\pi\pi$ cross section is negligibly weak as
long as the typical momentum is in range of validity of the
low energy theorem \cite{low}.
We approximate it by the threshold value (by doing so
we underestimate the effect).
According to the soft-pion theorem,
the model-independent low energy cross section is given by
\begin{equation}
\sigma(\pi^+\pi^-\rightarrow\pi^0\pi^0)= \frac{s}{32\pi f_{\pi}^4}
\left( 1-\frac{m_\pi^2}{s}\right) ^2,
\end{equation}
whose threshold value is
$\sigma_{\pi\pi}|_{\rm threshold}=9m_\pi^2/128\pi f_\pi^4$.
The integration over $t$ in (\ref{in}) can be done trivially,
\begin{equation}
{\cal F}(\infty) = \frac{n(0)\langle
v_{\rm rel}\sigma_{\pi\pi}\rangle R_D}{v_{\rm d}},
\end{equation}
where $n(0)=N/V(0)$ is the initial {\it total} pion density at the time
when the DCC decays.
Clearly, ${\cal F}(\infty)$ sensitively
depends on the average relative velocity
defined in a frame-independent way
$v_{\rm rel}={\sqrt{(p_1\cdot p_2)^2-m_\pi^4}}/{E_1E_2}$.
If the collision occurs
frequently during diffusion, the diffusion velocity would be much smaller
than the average pion velocity. In the other limit, if the
collision rarely occurs,
the diffusion takes place as pions stream freely.
In the latter case $v_{\rm d}$
is equal to
$\langle  |\vec{p}|\rangle/E$, and the average relative velocity is calculated
by integrating $v_{\rm rel}$  over all relative directions:
$\langle   v_{\rm rel}\rangle
\simeq \frac{4}{3}\langle  |\vec{p}|\rangle/E$ in the
non-relativistic limit
if all pions have approximately common $\langle  |\vec{p}|\rangle$.
Thus, we  obtain
\begin{equation}
{\cal F}(\infty)\simeq \frac{4}{3}\frac{R_D}{l_0},
\end{equation}
where the mean free path at $t=0$ is $l_0=1/(n(0)\sigma_{\pi\pi} )$.
The initial pion density is determined by the energy density of
the DCC
divided by the average pion energy, $n(0) = {m_{\pi}^2 f_{\pi}^2/m_{\pi}}$ in
the soft pion limit. Therefore the mean free path $l_0$ for the
conversion $\pi^+\pi^-\rightarrow\pi^0\pi^0$ at the threshold energy is given
by
 $l_0 \simeq  {128\pi f_\pi^2}/{9m_{\pi}^3}\simeq 27.7 \;{\rm fm}$,
where $f_{\pi}=93$ MeV.  It is
longer than the typical size of the DCC.  As the cloud of quanta diffuses, the
ratio of the mean free path to the cloud size increases even further.
Therefore,
for most DCC's, the diffusion of the pion cloud is closer to
a free streaming than to a
strongly interacting case.  We thus justify the  approximation in
calculating the diffusion
velocity $v_{\rm d}$ and  the average pion velocity.
For a DCC with a radius $R_D = 5 $fm,
${\cal F}(\infty)$ is about $0.24$.
If the DCC consists of only $\pi^0$'s, i.e.\
$f_0=1$, then $f(\infty)\simeq 80\%$, i.e.\ $20\%$ of original $\pi^0$'s are
converted into $\pi^\pm$'s!

We can also calculate the distribution $dP/df$ of the final neutral charge
fraction $f=N_0(\infty)/N$.
The classical field theory predicts
 ${dP}/{df_0} ={1}/{2\sqrt{f_0}}$
where $f_0=N_0(0)/N$ is the initial value at the time when the DCC decays.
The distribution in terms of $f$ is given by
\begin{equation}
\frac{dP}{df} = \frac{dP}{df_0}\cdot \frac{df_0}{df}
=\frac{1}{2\sqrt{f_0(f)}}\cdot
\frac{16e^{-{\cal F}}}{[(1+3e^{-{\cal F}})-3f(1-e^{-{\cal F}})]^2},
\end{equation}
where $df_0/df$ is computed from (\ref{f}).
The distribution $dP/df$ tends to concentrate more in the range around
$f=1/3$ and fall off
faster from one end to the other as $f$ increases, making the distinction from
non-DCC events a little less conspicuous.
In Fig.\ 1 we have plotted $dP/df$ for two values:
$R_D = 4$ fm  and $10$ fm.  The deformation is
not too severe to significantly affect the experimental test of (\ref{frac})
 for $R_D=4$ fm, but severe enough to make us worry for $R_D = 10$ fm.
For values
larger than $R_D = 10$ fm, the mean free path of collision becomes comparable
or even shorter than the DCC size.  Our assumption that
$\langle   v_{\rm rel}\rangle\simeq 4v_{\rm d}/3$ breaks down and
$v_{\rm d}$ becomes
slower than the
average pion velocity.  In other words, pions emitted  stick together through
final-state interactions.  When this happens,
the dilution factor ${\cal F}(\infty)$
rises quickly with $ R_D$, making the experimental test of the
classical theory prediction difficult.

\section*{Acknowledgements}
We would like to thank X.-N.\ Wang, K.\ Rajagopal and M.\ Bander
 for discussions.
This work was supported in part by the Director, Office of
Energy Research, Office of High Energy and Nuclear Physics, Division of
High Energy Physics of the U.S. Department of Energy under Contract
DE-AC03-76SF00098 and in part by the National Science Foundation under
grant PHY-90-21139.
Z.H.\ is also supported by the Natural Sciences and
Engineering Research Council of Canada.

\newpage
\centerline{\bf Figure Captions}
\vskip 15pt
\begin{description}
\item[Fig. 1] Comparison of the probability distributions of
the neutral pion fraction
$f=N_{\pi^0}/N_{\rm total}$ for different sizes of DCC.
\end{description}
\end{document}